\documentclass[book]{elsbook}

\usepackage{amsmath,amssymb,amsfonts,amsthm,makeidx,graphicx,colortbl}
\usepackage{txfonts}
\usepackage{helvet}

\begin{document}

\begin{frontmatter}

\chapter[The dynamics of rings around Centaurs and Trans-Neptunian Objects]{%
The dynamics of rings around Centaurs and Trans-Neptunian Objects
}%
\label{chap_rings}

\author*[1]{Bruno Sicardy} 
\author[2]{Stefan Renner}%
\author[3]{Rodrigo Leiva }%
\author[1]{Fran\c{c}oise Roques}%
\author[4,5]{Maryame El Moutamid}%
\author[6]{Pablo Santos-Sanz}%
\author[1]{Josselin Desmars}%

\address[1]{%
\orgname{Observatoire de Paris, PSL Research University, CNRS, Sorbonne Universit\'e, Univ. Paris Diderot, Sorbonne Paris Cit\'e}, 
\orgdiv{LESIA}, 
\orgaddress{5 place Jules Janssen, 92195 Meudon, France}
}%
\address[2]{%
\orgname{Observatoire de Paris, CNRS UMR 8028, Universit\'e de Lille, Observatoire de Lille}, 
\orgdiv{IMCCE}, 
\orgaddress{1, impasse de l'Observatoire, F-59000 Lille, France}
}%
\address[3]{%
\orgname{Southwest Research Institute}, 
\orgdiv{Dept. of Space Studies}, 
\orgaddress{1050 Walnut Street, Suite 300, Boulder, CO 80302, USA}
}%
\address[4]{%
\orgname{Cornell University}, 
\orgdiv{Center for Astrophysics and Planetary Science}, 
\orgaddress{Ithaca, NY 14853, USA}
}%
\address[5]{%
\orgname{Cornell University}, 
\orgdiv{Carl Sagan Institute}, 
\orgaddress{Ithaca, NY 14853, USA}
}%
\address[6]{%
\orgname{Instituto de Astrof\'{\i}sica de Andaluc\'{\i}a (CSIC)}, 
\orgdiv{CSIC}, 
\orgaddress{Glorieta de la Astronom\'{\i}a S/N, 18008-Granada, Spain}
}%

\address*[1]{Corresponding: \email{bruno.sicardy@obspm.fr}} 

\titlemark{Part Title}
\chaptermark{The dynamics of rings around Centaurs and Trans-Neptunian Objects}

\MaxmMiniTocnum{A.1}{3.3.3}{}


\makechaptertitle

\begin{abstract}[Abstract]
Since 2013, dense and narrow rings are known 
around the small Centaur object Chariklo \cite{bra14} 
and the dwarf planet Haumea \cite{ort17}.
Dense material has also been detected been around the Centaur Chiron,
although its nature is debated \cite{ort15,rup15}. 
This is the first time ever that rings are observed elsewhere than around the giant planets,   
suggesting that those features are more common than previously thought.
The origins of those rings remain unclear. 
In particular, it is not known if the same generic process can explain the presence of material 
around Chariklo, Chiron, Haumea, or if each object has a very different history.
Nonetheless, a specific aspect of small bodies is that they may possess  
a non-axisymmetric shape (topographic features and/or elongation) 
that are essentially absent in giant planets.
This creates strong resonances between the spin rate of the object and
the mean motion of ring particles.
In particular, Lindblad-type resonances tend to clear the region around the 
corotation (or synchronous) orbit, 
where the particles orbital period matches that of the body.
Whatever the origin of the ring is, 
modest topographic features or elongations of Chariklo and Haumea
explain why their rings should be found beyond the outermost 1/2 resonance,
where the particles complete one revolution while the body completes 
two rotations. 
Comparison of the resonant locations relative to the Roche limit of the body
shows that fast rotators are favored for being surrounded by rings. 
We discuss in more details the phase portraits of the 1/2 and 1/3 resonances, 
and the consequences of a ring presence on satellite formation.
\end{abstract}

\begin{keywords}[Keywords:]
Trans-Neptunian Objects \sep 
Centaurs \sep 
Ring dynamics \sep 
Resonances
\end{keywords}

\end{frontmatter}%

\section{Introduction}
\label{chap_rings:intro}

Trans-Neptunian Objects (TNOs) are very remote and faint objects.
They orbit at some 30 to 48 au from the Sun 
and have magnitudes typically fainter than 18, thus requiring large telescopes.
They subtend small angular diameters, e.g. at most 100 milli-arcsec (mas) or so for the largest one, Pluto,
and less than 35 mas for other large ones like Eris and Makemake.
Centaur objects, that are thought to be dynamically related to TNOS, orbit closer to us
between Jupiter and Saturn, but they are small (300~km at most) and thus faint too and no more
than 25~mas in angular size as seen from Earth. 
Classical imaging, either using adaptive optics or space telescopes, cannot 
provide any accurate details about their sizes, shapes and surroundings.

In that context, stellar occultations are a precious tool, in particular to detect rings 
around small bodies, see also Chapter~22.
Out of about 25 TNOs or Centaurs probed up to now using occultations, 
two proved to be surrounded by rings,
providing a rather high frequency of about 8\% for the presence of rings around those remote bodies.
First in 2013, when two narrow and dense rings were discovered around Chariklo \cite{bra14,sic18a},
and then in 2017, when a ring was spotted around the dwarf planet Haumea \cite{ort17}.
Physical parameters of those rings are listed in \weblink{Table}{chap_rings:tab_ring_parameters}.
Note that dense and narrow material has also been detected around the second largest known Centaur, Chiron.
It could be a ring system alike that of Chariklo \cite{ort15}
(in which case the frequency of rings would be about 12\%),
or a bound cometary dust shell \cite{rup15}.

This detection rate is statistically high, an augury for more ring detections around other bodies. 
Those new rings were the first detected elsewhere than around the four giant planets,
and they may outnumber them soon. 
Moreover, the occultation technique now greatly benefits from the Gaia catalogues \citep{gai16,gai18}.
Combining astrometric measurements of the occulting bodies, past stellar occultations
and the Gaia catalogues, accuracies at the level of 10~mas (corresponding to some 200~km at the object)
can now be reached for predicting these events,
and even better for some bodies like Chariklo or Pluto, for which several occultations have been detected \citep{des17,des18}.

Those recent (and possibly future) detections  launched a new wave of interests on ring studies.
First because they can tell us something about the history of the object they surround,
and second because they reside in a dynamical environment that is specific to small bodies,
thus revealing some hitherto unstudied mechanisms.

\begin{table}[!t]
\TBL{\caption{Chariklo and Haumea's rings: physical parameters}
\label{chap_rings:tab_ring_parameters}}
{%
\tabcolsep15pt %
\renewcommand{\arraystretch}{1.3} 
\begin{tabular*}{\textwidth}{lll}
\hline 
Parameters & Chariklo & Haumea \\
\hline 
\multicolumn{3}{c}{Main body} \\
\hline 
Rotation period, $T_{\rm rot}$ (h) \cite{for14,lel10} &     7.004    &  3.915341  \\
\hline
Mass $M$ (kg) \cite{lei17,rag09} & 6.3$\times$10$^{18}$ &      4.006$\times$10$^{21}$  \\
\hline
Rotational parameter $q$$^{(a)}$ & 0.226 &  0.268 \\
\hline
Semi-axes $A\times B \times C$ (km) \cite{lei17,ort17} & 157$\times$139$\times$86$^{(b)}$ & 1161$\times$852$\times$513 \\
\hline
Reference radius $R$$^{(c)}$ (km) & 115 & 712 \\
\hline
Elongation parameter$^{(d)}$   $\epsilon$ &  0.20 & 0.61 \\  
\hline
Oblateness parameter$^{(d)}$ $f$ &  0.55 &  0.76 \\ 
\hline 
\multicolumn{3}{c}{Rings} \\
\hline 
Pole position (deg)   $\alpha_{\rm p}$   &  151.25$\pm$0.50 & 285.1$\pm$0.5$^{(e)}$ \\
(equatorial J2000)  $\delta_{\rm p}$     &   41.48$\pm$0.22  & -10.6$\pm$1.2 \\ 

\hline
Width (km) \cite{ber17,ort17} & C1R: $\sim$5-7 & $\sim$70 \\
                                              & C2R: $\sim$1-3 & \\
\hline
Apparent transmission$^{(f)}$ \cite{ber17,sic18a,ort17} & C1R: $\sim$0.5 &  $\sim$0.5 \\
                                                                                        & C2R: $\sim$0.9 & \\
\hline
Visible reflectivity $(I/F)_{\rm V}$  \cite{duf14}               &  0.07 & unknown \\                                                                   
\hline
Composition$^{(g)}$ \cite{duf14,ort17} & 20\% water ice & unknown \\
                                                            & 40-70\% silicates & \\
                                                            & 10-30\% tholins & \\
                                                            & some amorphous carbon & \\
\hline
Orbital radii (km) \cite{bra14,ort17} & 390 (CR1) & 2287 \\
                                                        & 405 (C2R) &          \\                                                     
\hline
Corotation radius $a_{\rm cor}$ (km)       & 189 & 1104 \\
Outer 2/4 resonance radius$^{(h)}$ (km) & 300 & 1752 \\
Outer 2/6 resonance radius$^{(h)}$ (km) & 392 & 2285 \\
\hline
Classical Roche limit $a_{\rm Roche}$$^{(i)}$ (km) & $\sim$280 & $\sim$2420 \\
\hline 
\end{tabular*}
}%
{%
\begin{tablenotes}
\footnotetext{
Notes.
$^{(a)}$\eqweblink{Eq.}{chap_rings:eq_q}.
$^{(b)}$Assuming a Jacobi equilibrium shape. Other solutions are possible \cite{lei17}.
$^{(c)}$\eqweblink{Eq.}{chap_rings:eq_R_appen}.
$^{(d)}$\eqweblink{Eqs.}{chap_rings:eq_epsilon_f_appen}.
$^{(e)}$The ring is coplanar with the orbit Haumea's main satellite Hi'iaka \cite{ort17}.
$^{(f)}$Fraction of light transmitted by the rings, as seen in the plane of the sky at present epoch. 
$^{(g)}$Still debated, see text.
$^{(h)}$Using \eqweblink{Eqs.}{chap_rings:eq_reson} and \eqweblink{}{chap_rings:eq_n_kappa_approx}.
$^{(i)}$Using \eqweblink{Eq.}{chap_rings:eq_roche} and assuming fluid particles ($\gamma$=0.85)
with icy composition ($\rho'$=1000 kg m$^{-3}$). This is probably unrealistic, see text.
}
\end{tablenotes}
}%
\end{table}

\section{Rings around irregular bodies}
\label{chap_rings:rings_gen}

Currently, little is known about the nature, origin and evolution of rings around small bodies like Chariklo or Haumea \cite{sic18a}.
Note that 
Haumea's ring composition is still unknown \cite{ort17}, 
while silicates, tholins and water ice may be present in Chariklo's rings 
(\cite{duf14} and \weblink{Table}{chap_rings:tab_ring_parameters}).
However, this latter point is debated because the suspected water ice 
feature in Chariklo's ring spectrum presented in \cite{duf14} 
is close to a gap in the spectral coverage of the instrument.
Also, it could be that the water ice feature is caused by Chariklo itself, and
that the rings are dark (like those of Uranus \cite{lei17}) and spectroscopically featureless.
The importance of water ice for forming rings has been suggested,
predicting ring existence in the heliocentric range 8-20 au \cite{hed15}.
This is at odd with the discovery of Haumea's ring, currently at 50~au from the Sun,
and with perihelion distance larger than 35~ua. 
In any case, this opens the interesting question of whether rings should be composed
of water ice (and why), a possibility that would favor ring presences in the TNO realm 
rather than in the asteroid Main Belt.

Those objects are angularly so small that detection of putative shepherd satellites near the rings 
is extremely challenging \cite{sic15}.
The chaotic nature of Chariklo's orbit suggests a scenario where that body has close encounter(s)
with a giant planet, thus triggering ring formation through tidal disruption \cite{hyo16}.
However, the probability of such encounters being small \cite{ara16,woo17}, 
this route to ring formation seems unlikely.
Satellite orbital evolution followed by disruption has also been invoked to form Chariklo's rings, 
as well as a possible impact onto the body by an external body \cite{mel17}.
But again such events are rare, and the satellite tidal disruption scenario should be
supported by more observational evidences, like the presence of a retinue of small satellites 
around Chariklo. As mentioned earlier, this is very challenging.
Concerning Haumea, a violent collisional history \cite{bro07} 
plus the coplanarity of the ring orbit with that of its main satellite Hi'iaka orbit
suggest co-genetic origins.

More dynamical works sought to constrain the mass of Chariklo's main ring C1R, 
using the fact that it may be eccentric, and should be maintained as such
through self-gravity, like those of Uranus \cite{pan16}. 
However, the eccentricity of CR1 is not well constrained \cite{ber17},
especially because the ring precession rate is not known \cite{sic18a}.
Thus, the similarity between Chariklo and Uranus' rings orbital structure has yet to be proved,
and inferences on the ring mass should be taken with caution.

Numerical studies using collisional codes have proved useful to describe the 
local behavior of Chariklo's rings.
But one of these codes does not include the non-spherical shape of Chariklo \cite{mic17}, 
while the other one does consider it, but without rotation \cite{gup18}, 
an important ingredient of ring dynamics, see below.

Considering the small amount of information currently available on rings 
around small objects, we focus here on some more basic and generic processes.
Let us first recall that a collisional disk around a central body dissipates energy,
while conserving its total angular momentum \textbf{H}. 
Assuming that the body rotates around its principal axis of largest inertia (i.e. that it does not wobble), 
and after averaging its potential $U(\textbf{r})$ over the azimuth, 
one obtains that the average angular momentum $<$\textbf{H}$>_z$ of a particle 
projected on the rotation $z$-axis of the body is conserved.
As a consequence, the system tends to flatten in the plane perpendicular to $<$\textbf{H}$>_z$
(called the equatorial plane herein), minimizing its energy while conserving $<$\textbf{H}$>_z$. 
The competing effects of pressure (stemming from the particle velocity dispersion $c$),
rotation and self-gravity lead to an equilibrium where the local Toomre's parameter $Q$
is a few times unity:
\begin{align}
\label{chap_rings:eq_toomre}
Q= \frac{c\kappa}{\pi G\Sigma_0} > \sim 1,
\end{align}
where $\kappa$ is the particle radial epicyclic frequency,
$G$ is the gravitational constant and $\Sigma_0$ is the disk surface density.
Let us consider a broad disk 
of radius $r$, vertical thickness $h$ and total mass $M_{\rm D}$ surrounding the body.
Noting that 
$Q \sim 1$,
$GM \sim r^3 \kappa^2$, 
$h \sim c/\kappa$ and
$M_{\rm D} \sim \pi r^2 \Sigma_0$, we obtain
\begin{align}
\label{chap_rings:eq_thickness}
\frac{h}{r} \sim \frac{M_{\rm D}}{M}.
\end{align}
Only crude mass estimations for the current masses $M_{\rm r}$ of Chariklo and Haumea's rings are possible.
Since the orbital periods of ring particles (some 10-20~h) are of same order as for Saturn's ring A,
the frequencies $\kappa$ are similar. 
Consequently, the local kinematic conditions in all those rings are expected to be comparable \cite{sic18a}.
This suggests that Chariklo's ring C1R, Haumea's ring and Saturn's ring A
share the same surface density, some $\Sigma_0 \sim$~500-1000~kg~m$^{-2}$ \cite{col09}. 
\weblink{Table}{chap_rings:tab_ring_parameters} 
then yields a ring mass estimate $M_{\rm r} \sim 10^{13}$~kg for C1R,
equivalent to an icy body of radius $\sim$1~km,
corresponding to very small fractions of Chariklo's mass, 
$M_{\rm r}/M_{\rm C} \sim 10^{-6}$
and angular momentum, $H_{\rm r}/H_{\rm C} \sim 10^{-5}$.
The same exercise for Haumea provides 
$M_{\rm r}/M_{\rm H} \sim 10^{-7}$ (equivalent to an icy body of radius $\sim$5~km) and
$H_{\rm r}/H_{\rm H} \sim 10^{-6}$.

Noting that broad disks around Chariklo's and  Haumea
would have masses about 100 times that of the current rings, 
we have from \eqweblink{Eq.}{chap_rings:eq_thickness}
$h/r \sim  M_{\rm D}/M \sim 10^{-5}-10^{-4}$, 
i.e. $h$ of some 10 meters, corresponding to a very flat disk. 
To keep the problem at its simplest, 
we will assume the body is not wobbling as mentioned earlier, 
that any departure of the body from axisymmetry is purely sectoral, 
i.e. depends only on longitude, not latitude. 

In that context, small objects like Chariklo or Haumea 
appear quite different from the ring-bearing giant planets, size put apart.
In fact, they may support much larger 
non-axisymmetric terms in their gravitational potentials.
For instance, a topographic feature of height or depth $z$ represents
a mass anomaly $\mu \sim (z/2R)^3$ relative to the body.
Considering 
$R=130$~km, typical of Chariklo, and 
$z\sim 10$~km (not excluded for such a body \cite{lei17}), 
we obtain $\mu \sim 10^{-4}$. 
This is comparable to Titan's mass relative to Saturn. 
Such a body placed on Saturn's equator would have devastating effects on the rings.
In fact, the saturnian satellite Janus is able to sharply truncate the outer edge of Saturn's ring A
with a mere $\mu \sim 3 \times 10^{-9}$, while mass anomalies inside the planet have at most
$\mu \sim 10^{-12}$ \cite{hed14}.
The differences between small bodies and giant planets is exacerbated 
when their shapes are considered. 
For instance Haumea is extremely elongated 
(elongation $\epsilon \sim 0.6$, see \weblink{Table}{chap_rings:tab_ring_parameters} and
\eqweblink{Eq.}{chap_rings:eq_pot_ell_appen}),
as is possibly Chariklo too ($\epsilon \sim 0.2$).
The mass anomalies stored in the  bulges of those bodies are of order $\epsilon$,
which still increases the difference between the essentially axisymmetric giant planets
and small bodies.

We focus here on the coupling between the central body and a surrounding collisional disk.
We will not discuss the very origin of this disk, as discussed before. 
We feel that a good understanding of the ring dynamics around a non-spherical body
is a necessary preliminary step before investigating further the origin of those rings.

\section{Potential of a non-axisymmetric body}
\label{chap_rings:potential}

As mentioned earlier, the body is assumed to rotate without wobbling.
Consequently,  each of its volume elements executes a circular motion at constant angular speed,
noted $\Omega$ herein ($T_{\rm rot}= 2\pi/\Omega$ being the rotation period).
The problem is simplified further by assuming that departure from sphericity is described 
either by a mass anomaly (under the form of a topographic feature) 
lying on the equator of a homogeneous sphere of mass $M$ and radius $R$, 
or by a homogeneous ellipsoid of principal semi-axes $A>B>C$ rotating around $C$.
In that context, a collisional disk, once settled in the equatorial plane of the body, 
does not feel any perpendicular force that would excite off-plane motions.

The vector \textbf{r} denotes the position vector of the particle (counted from the center of the body),
with $r=||\textbf{r}||$.
The angle $L$ is  the true longitude of the particle (counted from an arbitrary origin) and
$L'=  \Omega t$ measures the orientation of the body
through its mass anomaly or its longest axis $A$, $t$ being the time.
By symmetry, the expansion of the potential then contains only cosine terms, with the generic form
\begin{align}
\label{chap_rings:eq_pot_generic}
U(\textbf{r}) = \sum_{m=-\infty}^{+\infty} U_{m}(r) \cos\left(m\theta \right),
\end{align}
where $\theta= L - L'$. The integer $m$ is called the azimuthal number.

In the mass anomaly case, 
$U(\textbf{r})$ can be split into a keplerian term, $-GM/R$, and 
terms depending on $\mu$ (see \cite{sic18b} for details), so that
\begin{align}
\label{chap_rings:eq_pot_ma}
U(\textbf{r}) = 
-\frac{GM}{r} - \frac{GM}{2R}\mu
\left\{\left[\sum_{m=-\infty}^{+\infty} b_{1/2}^{(m)}(r/R) \cos (m\theta)\right]  
- 2q \left(\frac{r}{R}\right) \cos(\theta) \right\},
\end{align}
where $G$ is the constant of gravitation and $b_{1/2}^{(m)}(r/R)$ are the classical Laplace coefficients.
The dimensionless coefficient 
\begin{align}
\label{chap_rings:eq_q}
q= \frac{R^3 \Omega^2}{GM}
\end{align}
measures the rotational stability of the body, so that $q<1$ for bodies of interest.

In the case of a homogeneous ellipsoid, 
\textit{only even values of $m$ are allowed}, 
ensuring the invariance of the potential under a rotation of $\pi$ radians.
As seen later, this will affect the orders of the resonances. 
Posing $m=2p$, we have 
\begin{align}
\label{chap_rings:eq_pot_ell_generic}
U(\textbf{r}) = \sum_{p=-\infty}^{+\infty} U_{2p}(r) \cos\left(2p\theta \right).
\end{align}
The calculation of $U_{2p}(r)$ is detailed in 
\weblink{Appendix}{chap_rings:appendix1}.
Defining the sequence $S_{|p|}$ recursively as 
\begin{align}
\label{chap_rings:eq_Sp}
S_{|p|+1} = 2
\frac{(|p|+1/4)(|p|+3/4)}{(|p|+1)(|p|+5/2)}
\times S_{|p|}
{\rm~~with~~}
S_0 = 1,
\end{align}
and turning back to $m=2p$, we obtain at lowest order in the elongation $\epsilon$ 
\begin{align}
\label{chap_rings:eq_pot_ell_approx}
U(\textbf{r}) = -\frac{GM}{r} 
\sum_{m=-\infty}^{+\infty} 
\left( \frac{R}{r} \right)^{|m|} 
S_{|m/2|} \epsilon^{|m/2|}
\cos\left(m \theta \right) ~~(m{\rm~even}),
\end{align}
\eqweblink{Eqs.}{chap_rings:eq_pot_ma} and \eqweblink{}{chap_rings:eq_pot_ell_approx}
define the potentials caused by a mass anomaly and by a homogeneous ellipsoid, respectively.
Note that $U_m(r)$ has an exponential behavior with $m$, 
i.e. $|U_m(r)| \propto K^{|m|}$, where $K<1$ is a constant
depending on the problem under study \cite{sic18b}.
Thus, the strengths of the resonances rapidly decrease as $|m|$ increases.

\section{Resonances around non-axisymmetric bodies}
\label{chap_rings:sec_reson_gen}

A particle moving in the equatorial plane of a body has two degrees of freedom.
They are associated with two fundamental frequencies: 
the mean motion $n$ and the radial epicyclic frequency $\kappa$,
whose calculations are outlined in \weblink{Appendix}{chap_rings:appendix2}.
However, the non-axisymmetric terms $\cos(m \theta)$ ($m \neq 1$) in \eqweblink{Eq.}{chap_rings:eq_pot_generic}
are time-dependent through $L'= \Omega t$,
introducing another degree of freedom.
For a rigid body rotating at constant angular speed $\Omega$ along a fixed axis, 
this dependence can be eliminated by using the variable $\theta= L-L'$ instead of $L$.
This is equivalent to writing the equations of motion in a frame corotating with the body.
The two fundamental frequencies of the autonomous system so obtained are now $n-\Omega$ and $\kappa$.
Two kinds of resonances happen, one being the corotation resonance
\begin{align}
\label{chap_rings:eq_corot}
n = \Omega,
\end{align}
also referred to as the synchronous orbit.

The other kind of resonances (called sectoral, see below)
occurs for 
\begin{align}
\label{chap_rings:eq_reson}
j\kappa= m(n-\Omega),
\end{align}
where $j$ and $m$ are integers.
Those two kinds of resonances are examined in turn.

\subsection{Corotation resonance}

The corotation region, where $n \sim \Omega$, can in principle host ring-arc material.
The particle motion is then best studied by re-expressing the potential
$U({\bf r})$ in a frame corotating with the body, i.e. 
\begin{align}
\label{chap_rings:eq_pot_corot}
V({\bf r}) = U({\bf r}) - \frac{\Omega^2 r^2}{2}.
\end{align}
Level curves of $V({\bf r})$ are shown in \weblink{Fig.}{chap_rings:fig_corotation}.
Note the presence of two stable points $C_2$ and $C_4$,
reminiscent of the Lagrange points $L_4$ and $L_5$, 
while $C_1$ and $C_3$ are reminiscent of the unstable Lagrange point $L_3$.

The points $C_2$ and $C_4$ being maxima of potential, 
they are unstable against dissipative collisions, even if dynamically stable. 
With the expected values of $\mu$ (some $10^{-5}$ for Chariklo), 
the librating (or ``tadpole") orbits around $C_2$ and $C_4$ are so narrow (a few km) that 
spreading time scales for removing the particles from that region 
is very short, a few years \cite{sic18b}.
An elongated body with $\epsilon=0.2$ creates much wider corotation zones,
and thus much longer spreading time scales (some $10^{4}$~yr).
However, another problem appears, as
the points $C_2$ and $C_4$ are linearly unstable for
\begin{align}
\label{chap_rings:eq_cor_stability}
\left(4\Omega^2 + V_{xx} + V_{yy}\right)^2 \leq V_{xx} V_{yy} - V^2_{xy},
\end{align}
where the indices $x$ and $y$ stand for partial derivatives \cite{mur99}.

Using \eqweblink{Eq.}{chap_rings:eq_pot_ma} and Chariklo's parameters,
this equation provides a critical value of about 0.04 for $\mu$,
above which $C_2$ and $C_4$ are linearly unstable.
This is unrealistically large, as this it requires a topographic feature of several tens of kilometers.  
Turning to the case of an elongated body, 
and limiting the expansion in \eqweblink{Eq.}{chap_rings:eq_pot_ell_approx} to the term $m=2$,
\eqweblink{Eq.}{chap_rings:eq_cor_stability} implies that $C_2$ and $C_4$ are unstable for
$\epsilon > \epsilon_{\rm crit} \sim 0.06/q^{2/3}$, 
i.e. $\epsilon_{\rm crit} \sim 0.16$ in the case of Chariklo ($q=0.226$).
Using the nominal value of Chariklo's elongation ($\epsilon=0.20$), this shows
that $C_2$ and $C_4$ are unstable for Chariklo, as illustrated in  \weblink{Fig.}{chap_rings:fig_corotation}.

In summary, the corotation region does not appear as a viable environment to host ring-arc material.

\begin{figure}[!t]
\FIG{
\includegraphics[width=0.9\textwidth]{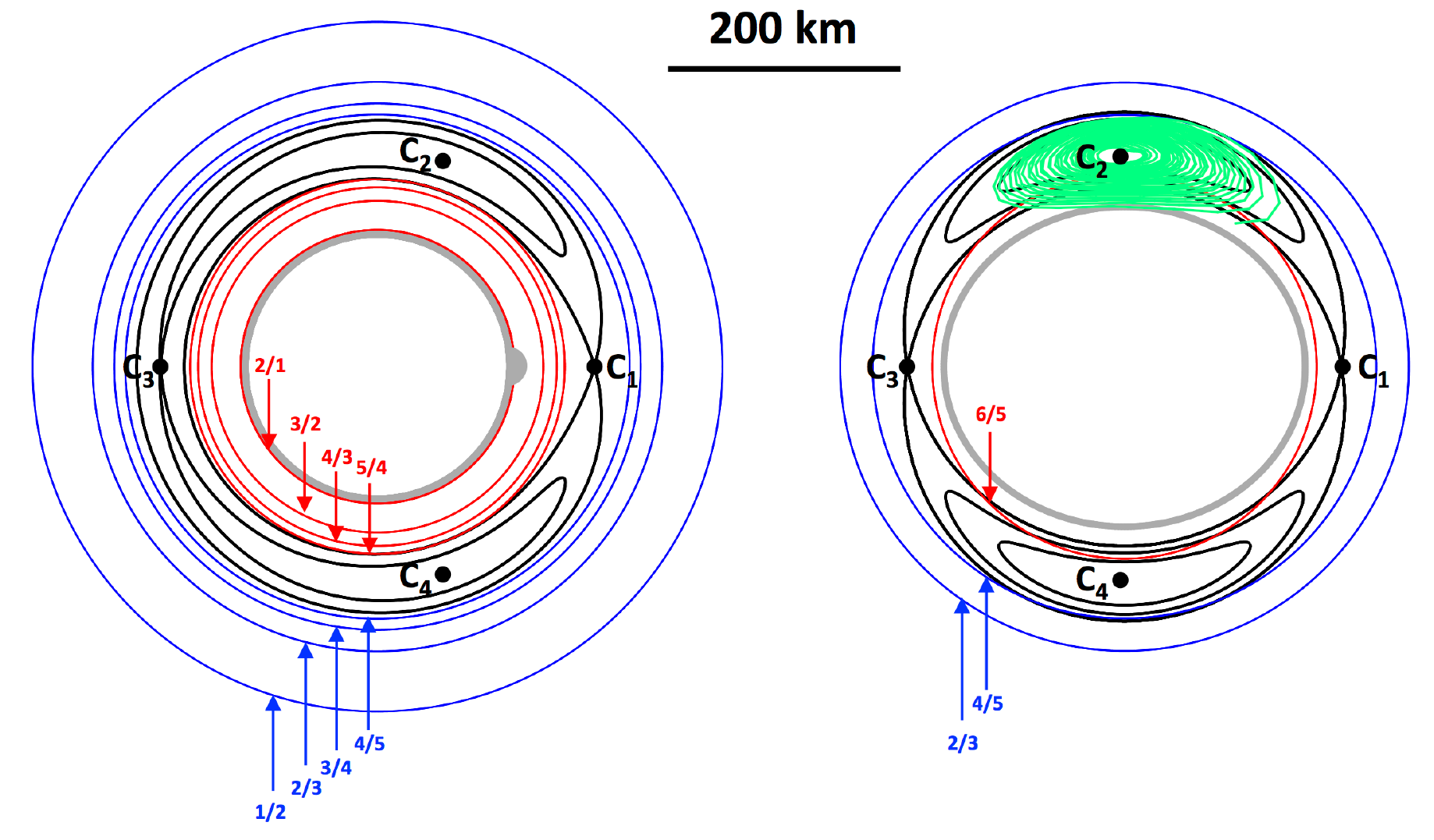}
}
{\caption{%
\title{%
Structure of corotation potential.
}%
Black lines: 
levels curves of the potential $V({\bf r})$ in the corotating frame (\eqweblink{Eq.}{chap_rings:eq_pot_corot}).
The parameters are those of Chariklo (\weblink{Table}{chap_rings:tab_ring_parameters}),
the bar at the top providing the distance scale. 
Left:
case of a mass anomaly on a spherical body ($R \sim130$~km) and $\mu=2.5 \times 10^{-2}$. 
This is much larger than Chariklo's topographic features (height $z \sim 5$~km,  $\mu \sim 10^{-5}$ \cite{lei17})
in order to better show the level curves around the stable points $C_2$ and $C_4$. 
Note the presence of two unstable points $C_1$ and $C_3$.
A few inner $m/(m-1)$ Lindblad Resonances radii
(LRs, see \weblink{Section}{chap_rings:sec_definition_LR})
are plotted in red ($m=2,...,5$),
while outer LRs are plotted in blue ($m=-1,...,-4$).
Right:
the same with an homogeneous ellipsoid with semi-axes 
$A \! \times \! B \! \times \! C$= 157$\times$139$\times$86~km and elongation $\epsilon=0.2$.
Only LRs with $m$ even are now present, with an example for $m=6$ (red) and $m=-2,-4$ (blue).
The green line shows the motion of a particle starting at $C_2$.
The trajectory is linearly unstable because $\epsilon$ is larger than the critical value $\epsilon_{\rm crit}=0.16$,
see text.
Adapted from \cite{sic18b}.
}%
\label{chap_rings:fig_corotation}}
\end{figure}

\subsection{Sectoral resonances}

The potential $U(\textbf{r})$ in \eqweblink{Eq.}{chap_rings:eq_pot_generic} 
can be expanded in a Fourier series containing harmonics $j\kappa - m(n-\Omega)$
of the two frequencies $\kappa$ and $n-\Omega$.
Without loss of generality, we can take $j \geq 0$, 
while $m$ is either positive or negative.
Other conventions are found in the literature, 
e.g. with $m$ positive and $j$ positive or negative. Our choice is motivated 
by the fact that it avoids the tedious use of the symbol $\pm$ in the equations.

The resonances studied here will be qualified as sectoral 
because they stem only from azimuth-dependent parts of the potential, 
with no dependence on latitude.
In other words, we do not consider here tesseral harmonics of the potential,
classically studied in the dynamics of artificial satellites of Earth, 
or associated with oscillation modes of giant planets \cite{mar93,mar14}.
We avoid the terminology ``spin-orbit resonances", 
that usually describes the response of a rotating secondary 
to the torque exerted by the primary, thus introducing confusions.
In fact, sectoral resonances are akin to mean motion resonances 
between a ring particle and an equatorial  satellite, 
the role of the satellite being played by the non-axisymmetric terms of the potential. 

The position vector of the particle, \textbf{r}, can be expressed in terms of 
the orbital elements $a$, $e$, $\lambda$ and $\varpi$, i.e. 
the semi-major axis, orbital eccentricity, mean longitude and pericentric longitude of the particle, 
respectively.
The relevant resonance angle associated with the resonance defined in 
\eqweblink{Eq.}{chap_rings:eq_reson} is
\begin{align}
\label{chap_rings:eq_phi_mj}
\phi_{m/(m-j)}= [m L' - (m-j)\lambda - j\varpi]/j,
\end{align}
recalling that $L'=\Omega t$.
Note the presence of the dividing factor $j$ in the expression above. 
This is needed to obtain the proper choice of canonical variables
in the Hamiltonian describing the resonance \cite{pea86}.

The expression of $\phi_{m/(m-j)}$ satisfies the d'Alembert's relation, i.e. it is invariant 
under a change of origin, since $m-(m-j)-j=0$.
Moreover, the d'Alembert's characteristics states that the amplitude 
of the term containing $\phi_{m/(m-j)}$ is of lowest order $e^{|j|}$ 
($=e^j$ since $j \geq 0$ by convention). 
For that reason, the integer $j$ is called herein the order of the resonance. 
In summary
\begin{align}
\label{chap_rings:eq_pot_ell_A_mj}
U(\textbf{r}) = U(a,e,\lambda,\varpi)= 
\sum_{m=-\infty}^{+\infty} \sum_{j=0}^{+\infty} A_{m,j}(a) e^j \cos\left(\phi_{m/(m-j)}\right),
\end{align}
where $A_{m,j}(a)$ depends on the problem under study.
The case $j=1$ (Lindblad Resonances) will be detailed in \weblink{Section}{chap_rings:sec_definition_LR}.

The condition\eqweblink{}{chap_rings:eq_reson} means that after $j$ synodic periods
$2\pi/(n-\Omega)$, the particle completes exactly $m$ radial oscillations,
which excites the particle orbital eccentricity \cite{elm14}.
Noting that the apsidal precession rate is $\dot{\varpi}= n -\kappa$,
\eqweblink{Eq.}{chap_rings:eq_reson} reads
\begin{align}
\label{chap_rings:eq_ratio_n_omega_exact}
\frac{n-\dot{\varpi}}{\Omega-\dot{\varpi}} = \frac{m}{m-j},
\end{align}
meaning that in a frame rotating at rate $\dot{\varpi}$, 
the particle completes $m$ revolutions while the body completes $m-j$ rotations.
Since usually $\dot{\varpi} \ll n, \Omega$, we have
\begin{align}
\label{chap_rings:eq_ratio_n_omega}
\frac{n}{\Omega} \sim \frac{m}{m-j},
\end{align}
loosely referred to as a ``$m/(m-j)$" resonance.

Without loss of generality, we may assume that $\Omega > 0$. 
If $n>0$ (and then also $\kappa > 0$), the particle has a prograde motion around the body.
Then $m>0$ implies that the resonance occurs inside the corotation radius, 
and will be qualified as an inner resonance. 
Conversely, $m<0$ will correspond to outer resonances. 
Note that we may have $n<0$ (with $\kappa < 0$) that describes a ``retrograde resonance",
in which the particle and the body move in opposite directions.
From $j > 0$, retrograde resonances always require $m > 0$.

\subsection{Resonance order}

We define the order of a resonance as the lowest power in eccentricity present
in the corresponding resonant term, e.g. the integer $j$ in \eqweblink{Eq.}{chap_rings:eq_pot_ell_A_mj}.
For a given $n/\Omega$, the order depends on the symmetry of the potential. 

For instance, due to its invariance under a $\pi$-rotation, 
an ellipsoid creates only terms with \textit{even} azimuthal numbers $m=2p$
(\eqweblink{Eq.}{chap_rings:eq_pot_ell_generic}), 
so that $n/\Omega= 2p/(2p-j)$.
Thus, to get $n/\Omega= 1/2$, one must take 
$m=2p=-2$ and $j=2$, corresponding to a second-order resonance. 
In the case of a mass anomaly, this ratio is obtained with $m=-1$ and $j=1$, 
corresponding to a first order resonance. 
More physically, one can see an ellipsoid
as a body with two opposite mass anomalies. The resonant term 
$\propto e \cos(2\lambda  - L' - \varpi)$          created by one bulge is canceled out  by the term 
$\propto e \cos[2\lambda  -  (L'+\pi) - \varpi]$ created by the other bulge.
This leaves the place for the second order term $\propto e^2 \cos(4\lambda  - 2L' - 2\varpi)$,
which is now invariant under $L' \rightarrow L' + \pi$.

For this reason, we use the notation 1/2 in the case of a mass anomaly, and
2/4 in the case of an elongated body, to enhance their respective second and fourth orders.
%
%
In the same vein, 
a mass anomaly creates a second order resonance $n/\Omega=1/3$ with $m=-1$,$j=2$,
while an elongated body creates a fourth order resonance $n/\Omega=2/6$ ($m=-2$,$j=4$),
that have respective potentials $\propto e^2$ and $\propto e^4$.

\section{Lindblad Resonances}

\subsection{Definition}
\label{chap_rings:sec_definition_LR}

The strongest resonances in \eqweblink{Eq.}{chap_rings:eq_reson} correspond to the lowest order, $j=1$.
They are called Lindblad Eccentric Resonances, or simply Lindblad Resonances 
(LRs herein\footnote{Lindblad Inclined Resonances are not considered here, as no off-plane forces are present.}).
Thus a LR corresponds to, recalling that $m$ can be positive or negative:
\begin{align}
\label{chap_rings:eq_LR}
\kappa= m(n-\Omega).
\end{align}
LRs have been extensively studied in the frame of galactic and ring dynamics.
One reason is that they are the strongest ones for a given $m$.
Another reason is that the resonant streamlines do not cross themselves,
contrarily to the cases $j > 1$, see below.
This avoids singularities in the hydrodynamical equations, and
allows an analytical treatment of the disk response to the resonant forcing.

A more general way to define a LR is in fact $\kappa= m(n-\Omega_p)$,
where $\Omega_p$ is the pattern speed of the forcing potential. 
In our case (a non-wobbling body), 
$\Omega_p=\Omega$ because each volume element of the body moves at constant speed along a circular path. 
This is not true for a perturbation caused by a satellite with orbital eccentricity $e_s$.
In that case, besides the fundamental pattern speed $n_s$ (the satellite mean motion), 
there is an infinity of other pattern speeds $n_s + \kappa_s/(m_s-1)$, 
where $\kappa_s$ is the satellite epicyclic frequency and $m_s$ is an integer.
The associated perturbing potential is then of order $e e_s^{m_s-1}$, and 
the order of the LR is said to be $m_s$, \textit{although it remains of first order in $e$}.
For instance, the second-order 5$:$3 Mimas LR in Saturn's A ring
is in fact a 4$:$3 first-order LR with one of the harmonics of Mimas' potential.

\subsection{Torques}
\label{chap_rings:sec_torques}

Each $m/(m-1)$ LR exerts a torque $\Gamma_m$ onto the disk \cite{sic18b}, 
\begin{align}
\label{chap_rings:eq_torque}
\Gamma_m= 
{\rm sign}(\Omega-n)
\left(\frac{4\pi^2 \Sigma_0}{3 n}\right) 
\frac{(GM)^2}{\Omega R^2}
{\cal A}_m^2,
\end{align}
where ${\cal A}_m$ is a dimensionless coefficient that measures the strength of the resonance,
see \weblink{Appendix}{chap_rings:appendix3}.
This coefficient depends on the coefficient $U_m(r)$ that appears in 
\eqweblink{Eqs.}{chap_rings:eq_pot_ma} and \eqweblink{}{chap_rings:eq_pot_ell_approx}
and its derivative, see \eqweblink{Eqs.}{chap_rings:eq_Am_ma} and \eqweblink{}{chap_rings:eq_Am_ell}.

The sign$(\Omega-n)$ factor shows that 
the material inside the corotation radius receives a negative torque and 
thus is pushed onto the body, 
while the material outside of the corotation is pushed outwards, 
and more precisely, outside the outermost 1/2 LR.
The time scales for these migrations are short, a few years only for bodies with the 
typical elongations $\epsilon$ of Chariklo and Haumea (\weblink{Table}{chap_rings:tab_ring_parameters}). 
Even a spherical body with a unique topographic feature of height 5~km sitting on its equator 
can clear the entire zone between its surface and the outer 1/2 LR in less than 1~Myr (\cite{sic18b}).

This model shows that rings can exist around a non-axisymmetric body 
only if the latter rotates fast enough. 
This is because the 1/2 LR at radius $a_{1/2}$ must lie inside the Roche limit $a_{\rm Roche}$ of the body,
to prevent the ring from accreting into satellites. 
One can estimate
\begin{align}
\label{chap_rings:eq_roche}
a_{\rm Roche} \sim \left(\frac{3}{\gamma}\right)^{1/3} \left(\frac{M}{\rho'}\right)^{1/3},
\end{align}
where $\rho'$ is the density of the ring particles
and $\gamma$ is a factor describing the particle 
shape \cite{tis13}. From Kepler's third law, we have $a_{1/2}= 2^{2/3} a_{\rm cor}$, 
(where $a_{\rm cor}$ is the corotation radius) 
so that thus the condition $a_{1/2} < a_{\rm Roche}$ reads
\begin{align}
\label{chap_rings:eq_criterium_ring}
\gamma \rho' < \frac{3}{4} \frac{\Omega^2}{G}.
\end{align}
Although $\gamma$ is poorly known, a nominal value of $\gamma \sim 1.6$ 
can be used\footnote{The classical Roche limit is obtain with $\gamma = 0.85$
and particles with same density as the central body.},
representing particles filling their lemon-shaped Roche lobes (\cite{tis13}).
A value $\rho' \sim 450$~kg~m$^{-3}$ stems from the typical density of the
small saturnian satellites (\cite{tho10}), thus a good proxy of ring particles.
\eqweblink{Equation}{chap_rings:eq_criterium_ring} finally yields $T_{\rm rot} < \sim 7$~h,
a condition met by both Chariklo and Haumea (\weblink{Table}{chap_rings:tab_ring_parameters}).

\begin{table}[!t]
\TBL{\caption{Resonance orders around an elongated body}
\label{chap_rings:tab_reson_strength}}
{\tabcolsep15pt %
\renewcommand{\arraystretch}{1.3} 
\begin{tabular*}{\textwidth}{lllll}
\hline
Azimuthal number & $m=+2$ & $m=+4$ & $m=+6$ & $m=+8$  \\
\hline 
$j=1$ (Lindblad Resonances)   &  $e \epsilon$  [2/1]    & $e \epsilon^2$ [4/3]      & $e \epsilon^3$ [6/5]       & $e \epsilon^4$ [8/7]   \\
\hline
$j=2$           &  apsidal                               & $*\!*\!**$                                    & $e^2 \epsilon^3$ [6/4]   & $*\!*\!**$                                     \\
\hline
$j=3$           & $e^3 \epsilon$ [-2/1]     & $e^3 \epsilon^2$ [4/1]  & $*\!*\!**$                                     & $e^3 \epsilon^4$ [8/5]  \\
\hline
$j=4$ \dots  & $e^4 \epsilon$ [-1]        & apsidal                                    & $e^4 \epsilon^3$ [6/2] & $*\!*\!**$                                     \\
\hline
Azimuthal number  &  $m=-2$ & $m=-4$ & $m=-6$ & $m=-8$  \\
\hline
$j=1$ (Lindblad Resonances)    &  $e \epsilon$  [2/3]    & $e \epsilon^2$ [4/5]      & $e \epsilon^3$ [6/7]       & $e \epsilon^4$ [8/9]   \\
\hline
$j=2$           &  $e^2 \epsilon$ [2/4]  & $*\!*\!**$                                    & $e^2 \epsilon^3$ [6/8]   & $*\!*\!**$                                     \\
\hline
$j=3$           & $e^3 \epsilon$ [2/5]  & $e^3 \epsilon^2$ [4/7]  & $*\!*\!**$                                     & $e^3 \epsilon^4$ [8/11]  \\
\hline
$j=4$ \dots  &  $e^4 \epsilon$ [2/6]  & $*\!*\!**$                                    & $e^4 \epsilon^3$ [6/10] & $*\!*\!**$                                     \\
\hline
\end{tabular*}}%
{%
\begin{tablenotes}
\footnotetext{
Each entry shows the leading terms $e^j \epsilon^{|m/2|}$ of a few 
$n/\Omega \sim m/(m-j)$ sectoral resonances around an elongated body.
The fractions in brackets are the corresponding ratios $n/\Omega$.
The ratios that have absolute values larger (resp. smaller) than one 
correspond to inner (resp. outer) resonances.
The ratios that have positive (resp. negative) values correspond to prograde (resp. retrograde) resonances.
Because only the leading terms are considered here, 
some resonances are not replicated (symbols $*\!*\!**$).
For instance the $m=-4, j=2$ resonance 
(corresponding to the second order ($e^2 \epsilon^2$) resonance $n/\Omega=4/6$)
is already considered in the $m=-2, j=1$ case, 
corresponding to the 2/3 first order ($e \epsilon$) LR. 
}
\end{tablenotes}
}%
\end{table}
\section{Beyond the first order}
\label{chap_rings:higher_order}

\subsection{Streamline self-crossings}

A lowest order in eccentricity, 
a particle moves at distance $\rho = a \left[ 1 - e \cos (M)\right]$ from the body, 
where $M= \lambda - \varpi$ is the mean anomaly.
\eqweblink{Eq.}{chap_rings:eq_phi_mj} shows that $M= (m/j)\theta + \phi_{m/(m-j)}$,
where $\theta= L - L' \sim \lambda - L'$ is the longitude of the particle as observed
in a frame co-rotating with the body.
All the particles in a given resonant streamline share the same resonant angle $\phi_{m/(m-j)}$,
so that  its polar equation is
\begin{align}
\label{chap_rings:eq_streamline}
\rho (\theta)= 
a \left[ 1 - e \cos \left(\frac{m}{j} \theta + \phi_{m/(m-j)} \right) \right],
\end{align}
showing that $\phi_{m/(m-j)}$ defines the orientation of the streamline relative to the body.
To simplify the expressions used hereafter, 
we take without loss of generality $\phi_{m/(m-j)}=0$,
so that $\rho (\theta)= a [ 1 - e \cos(m\theta/j)]$,
noting that $e$ depends on $a$.

For a LR ($j=1$),
the polar equation $\rho (\theta)= a [ 1 - e \cos(m\theta)]$ shows that the resonance
forces a $m$-lobbed pattern on the streamline, 
that propagates as a $m$-armed spiral wave in the disk.
In this case, the function $\rho (\theta)$ is single-valued, 
and a given streamline does not have self-crossing points\footnote{
However, \textit{adjacent} streamlines may cross if the local gradient of eccentricity $de/da$ is large enough.}.
For $j > 1$, self-crossing usually occurs
because the same value of $\theta$ (mod $2\pi$) may correspond to different values of $\rho(\theta)$, 
providing a multivalued function. 
To see that, 
let us first divide $m$ and $j$ by their greatest common divider, 
\textit{so that $m$ and $j$ are now relatively prime}. 
Consider the sequence of $j$ angles
$\theta_0=\theta$, $\theta_1=\theta + 2\pi$,... $\theta_k= \theta + 2k\pi$,... ($0 \leq k < j$).
They correspond to the same angular position of a particle in the rotating frame. 
Since $m$ and $j$ are  relatively prime, $k(m/j)$ cannot be an integer for $0 < k < j$. 
Thus, all the angles $(m/j)\theta_k$, taken two by two, are different (mod $2\pi$), 
and so are in general the corresponding values of the radius $\rho (\theta_k)$, 
except at self-crossing points, see below. 
Consequently, the streamline exhibits $j$ distinct braids, each defined by a given $k=0,...j-1$.
Note that the streamline eventually closes onto itself for $k=j$, since $(m/j)\theta_j= (m/j)\theta_0 + 2m\pi$. 

\begin{figure}[!t]
\FIG{
\includegraphics[width=0.9\textwidth]{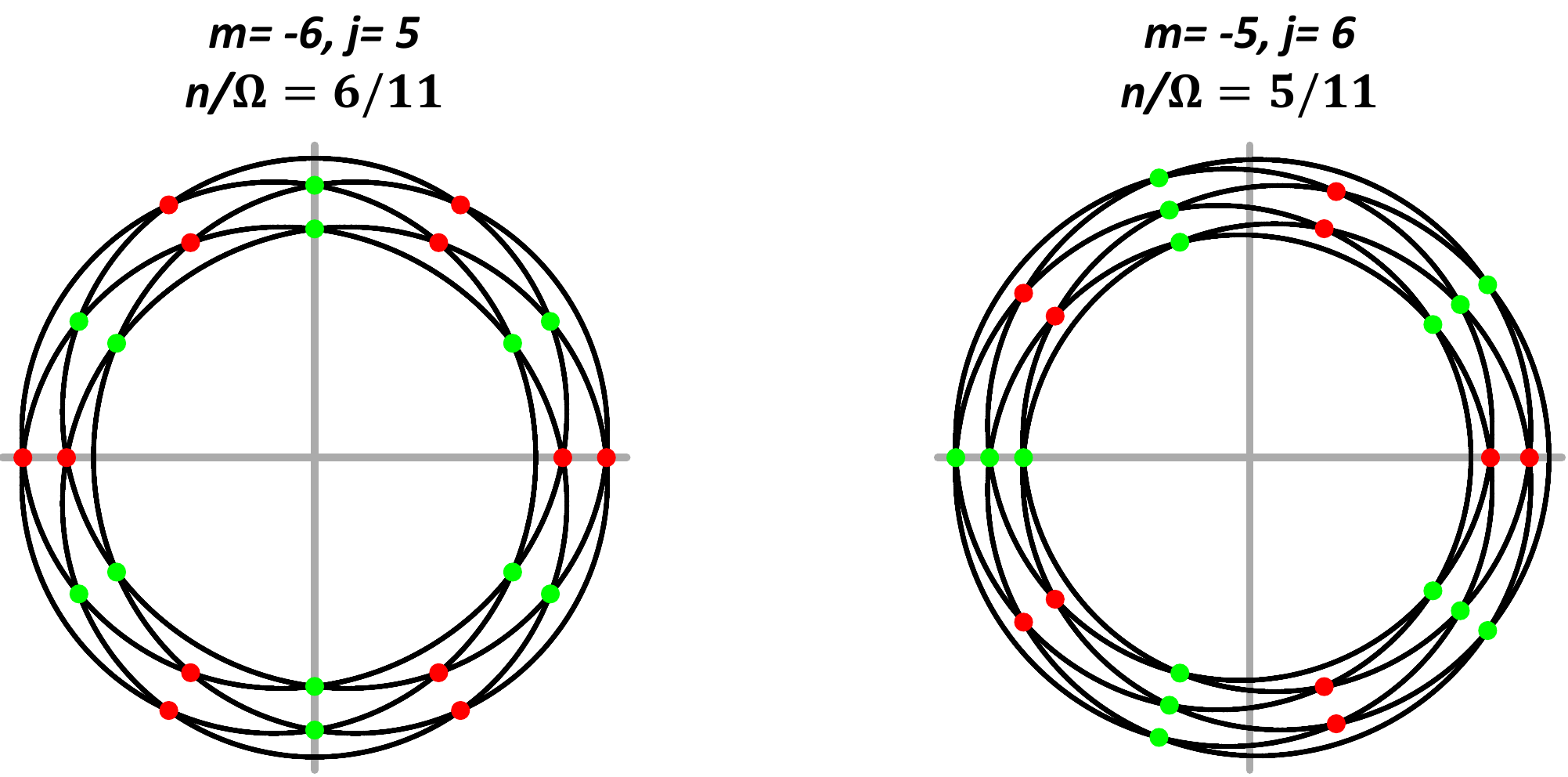}
}
{\caption{%
\title{%
Example of self-crossing resonant streamlines.
}%
Left: a periodic streamline with $m$=-6, $j$=5,
corresponding to $n/\Omega= m/(m-j)$=6/11, observed in a corotating frame.
Note the presence of $j$=5 braids and $|m|(j-1)$=24 self-crossing points.
The difference between the red and green points is discussed in \weblink{Fig.}{chap_rings:fig_circles_Mk}.
Right: the case $m$=-5, $j$=6 ($n/\Omega= 5/11$).
The streamline has now $j$=6 braids and $|m|(j-1)$=25 self-crossing points.
}%
\label{chap_rings:fig_streamline}}
\end{figure}

The existence of several braids means that the streamline must cross itself for $j > 1$,
see e.g. \weblink{Fig.}{chap_rings:fig_streamline}.
To count the number of self-crossing points, let us first show that the streamline 
is invariant by a rotation of $2\pi/m$ in the rotating frame.
Then it is enough to count the crossing points for $\theta \in [0,2\pi/|m|[$ and multiply the result by $|m|$.

The equation of the streamline, once rotated by $2\pi/m$, is
$\rho'(\theta)= \rho(\theta - 2\pi/m)= a \{1 - e \cos[(m\theta-2\pi)/j] \}$.
Thus $\rho'(\theta_k)= a \{1 - e \cos[(m\theta +(km-1)2\pi)/j] \}$.
From Bachet-B\'ezout theorem, and because $m$ and $j$ are relatively prime, 
there exists two integers $x$ and $y$ such that $mx + jy=1$.
Thus 
$\rho'(\theta_k)= 
a \{1 - e \cos[m(\theta+2(k-x)\pi)/j] \}=
a \{1 - e \cos[m(\theta+2k'\pi)/j] \}$
after posing $k'=k-x$.
Thus $\rho'(\theta_k)$ takes the same values as $\rho(\theta_k)$, 
except for a circular permutation on the indices $k=0,...j-1$,
showing the invariance of the streamline under a rotation of $2\pi/m$.
In other words, the streamline can be divided in identical sectors,
each of angular aperture $2\pi/|m|$
(e.g. $360/6=60$~deg in  \weblink{Fig.}{chap_rings:fig_streamline}, left panel,
and $360/5=72$~deg in the right panel).

Let us now consider $\theta$ varying in $[0,2\pi/m[$, 
corresponding to the first sector of the streamline.
Let us pose $M_k= m(\theta + 2k\pi)/j$, with $k=0,...,j-1$.
As $\theta$ goes from 0 to $2\pi/m$, 
$M_k$ goes from $2\pi(mk)/j$ to  $2\pi(mk)/j + 2\pi/j$
and $\rho(M_k)= a [1 - e \cos(M_k)]$ then describes one of the $j$ braids of the streamline.
The extremities of each angular interval  $[2\pi(mk)/j, 2\pi(mk)/j + 2\pi/j[$
can be distributed on the unit circle (\weblink{Fig.}{chap_rings:fig_circles_Mk}).
Because $m$ and $j$ are relatively prime, this defines exactly $j$ points on the circle
for $0 \leq k < j$.
The self-crossing points then correspond to those angles that have the same $\cos(M_k)$,
i.e. to the points that have the same projection onto the $x$-axis (\weblink{Fig.}{chap_rings:fig_circles_Mk}).
This occurs at $2(j-1)$ angular positions along the circle,
corresponding to $j-1$ self-crossing points as $\theta$ varies in $[0,2\pi/m[$, 
and thus a total of $|m|(j-1)$ self-crossing points in the entire streamline, 
see examples in \weblink{Fig.}{chap_rings:fig_streamline}.

\begin{figure}[!t]
\FIG{
\includegraphics[width=0.9\textwidth]{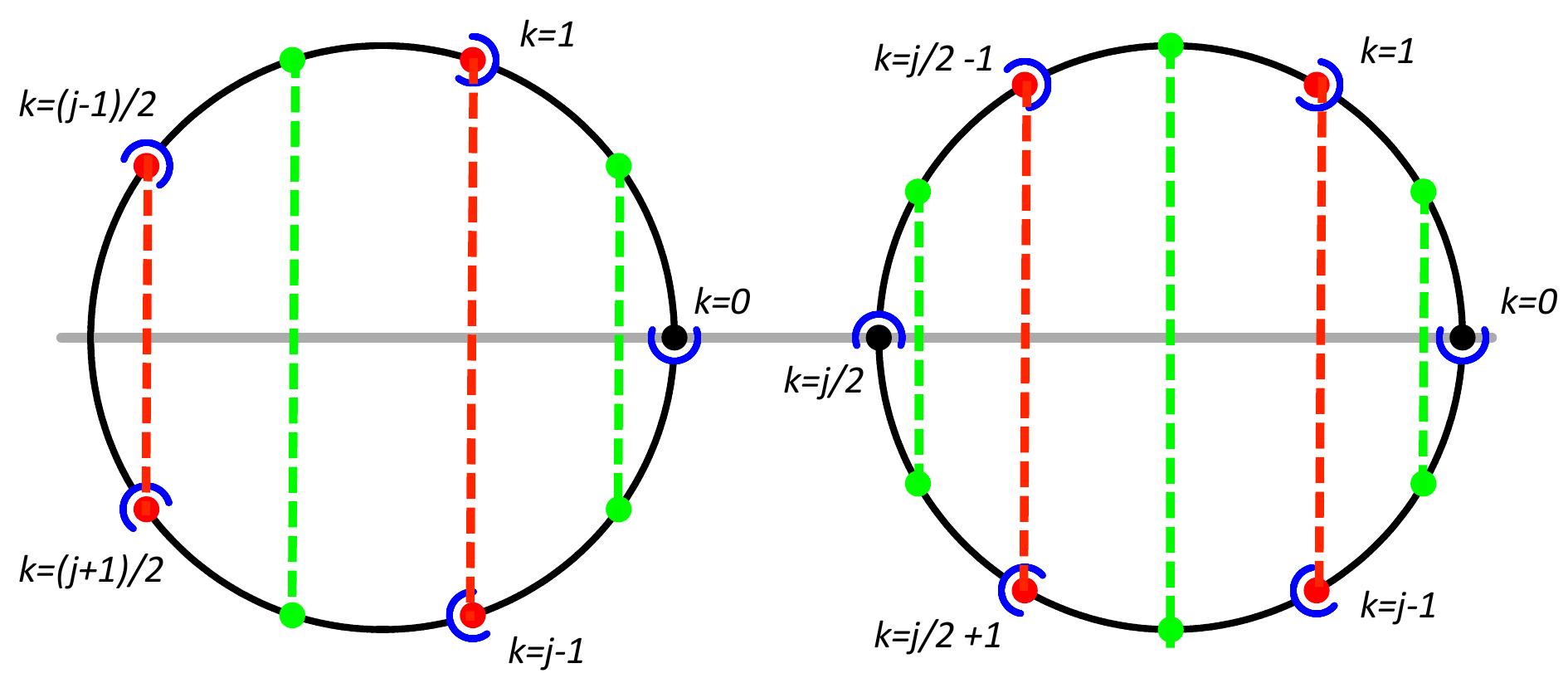}
}
{\caption{%
\title{%
Counting the self crossing points of a resonant streamline.
}%
Consider a $m/(m-j)$ resonance of order $j$ (with $m$ and $j$ relatively prime) and
divide the unit circle in $j$ identical arcs with opening angles $2\pi/j$.
The points with angular positions $M_k= m(\theta + 2k\pi/|m|)/j$ ($0 \leq k < j-1$)
then move simultaneously along each arc, from its closed extremity (dots)
to its open extremity (blue semi-circles), 
as $\theta$ varies in the interval $[0,2\pi/|m|[$.
Those points have a common projection onto the $x$-axis (gray line) for a given $\theta$
at the red or green points.
Left panel:  an example with $j$=5 odd.
There are $j$-1=4 red points and $j$-1=4 green points,
i.e. a total of $2j$-2=8 points that provides $j$-1=4 self-crossing points in the streamline,
defined by the 4 vertical dashed lines.
Right panel: the same, but with $j$=6 even.
There are now $j$-2=4 red points and $j$=6 green points, i.e. again a total of $2j$-2=10 points 
defining $j$-1=5 self-crossing points. 
It can be shown that this result is general, 
i.e. that the number of self crossing points is always $j$-1 
as $\theta$ moves in the sector $0 \leq \theta < 2\pi/|m|$. 
}%
\label{chap_rings:fig_circles_Mk}}
\end{figure}

Note that this result is valid only if $m$ and $j$ are relatively prime. 
However, as discussed in the previous subsection, it is desirable to
enhance the order of the resonance by using non relatively prime integers.
For instance the notation 2/6 resonance ($m=-2,j=4$) 
enhances its fourth order nature in the case of an elongated body,
but its irreductible version 1/3 ($m=-1,j=2$) must be used to calculate
the number of self-crossing points of the corresponding streamline, 
$|-1|\times(2-1)=1$.

In summary, a $m/(m-j)$ resonance is of order $j$, 
as long as $m$ and $j$ have not yet been reduced to their relatively prime version.
Once $m$ and $j$ are reduced to their relatively prime version, the resonant streamline has
\begin{itemize}
\item $|m|$ identical sectors, each of extension $(2\pi/|m|)$,
\item $j$ braids,
\item $|m|(j-1)$ self-crossing points.
\end{itemize}

Resonances are often noted $n/\Omega= p/q$ in the literature, instead of $m/(m-j)$.
Then it is of order $|p-q|$ before $p$ and $q$ have been reduced. 
After reduction to their relatively prime version,
the results above show that the resonant streamline has $|p|(|p-q|-1)$ self-crossing points.

\subsection{Phase portraits of 1/2 and 1/3 resonances}

We present here examples of phase portraits 
corresponding to outer sectoral resonances.
We focus on the 1/2 and 1/3 resonances caused by a mass anomaly, and
on the 2/4 and 2/6 sectoral resonances caused by an elongated body.
This stems from the fact that the 1/2 is the outermost LR where the classical torque formula
\eqweblink{Eq.}{chap_rings:sec_torques} applies, 
while the 2/4 resonance, although not a LR, has still a significant effect on the ring \cite{sic18b}.
Moreover, the 1/3 and 2/6 resonances are close to the locations of both Chariklo and Haumea's rings 
(\weblink{Table}{chap_rings:tab_ring_parameters}),
and thus deserve a special attention.

We do not care at this stage about the particular numerical values of $\mu$ or $\epsilon$, 
but rather about the topologies of the phase portraits and 
their consequences on a ring.
\begin{figure}[!t]
\FIG{
\includegraphics[width=0.9\textwidth]{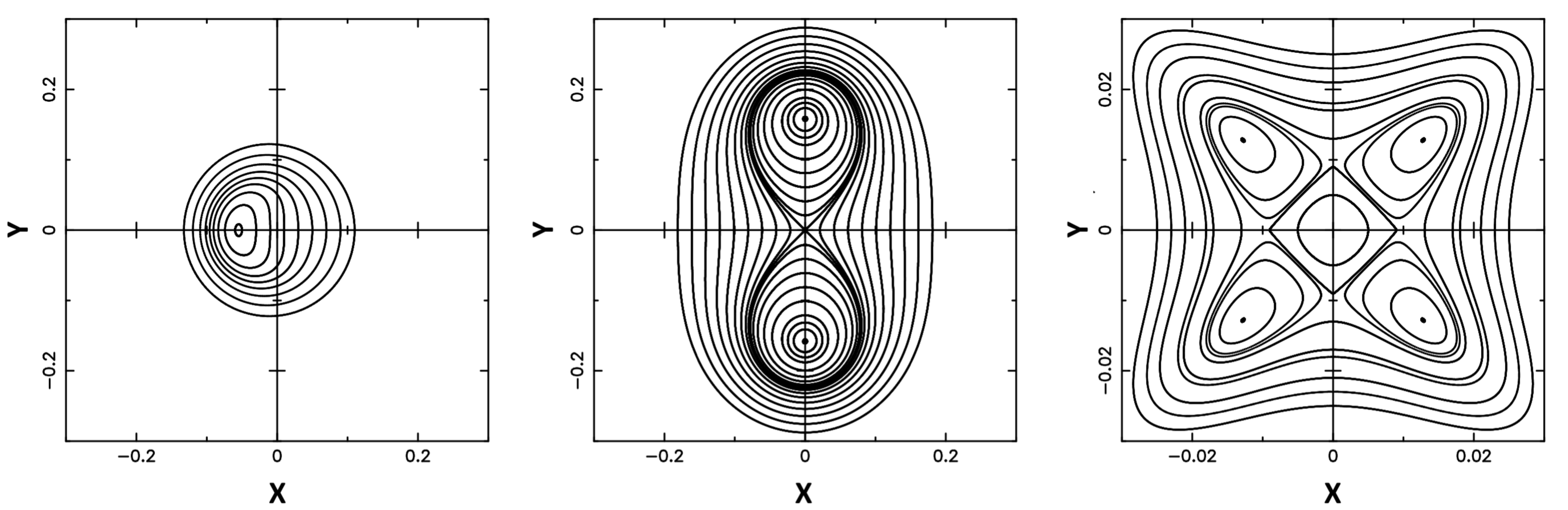}
}
{\caption{%
\title{Phase portraits of first, second and fourth order resonances.}
The trajectories of the eccentricity vector $(X,Y)= (e\cos\phi_{m/(m-j)},e\sin\phi_{m/(m-j)})$
are shown for various resonance orders. 
Left~-
Representative phase portrait of the first-order resonance 
$n/\Omega=1/2$, $\phi_{1/2}= 2\lambda - L' - \varpi$, caused by a mass anomaly.
Center~-
Phase portrait close to a second-order resonance, either caused by a mass anomaly with 
$n/\Omega=1/3$, $\phi_{1/3}= (3\lambda - L' - 2\varpi)/2$, 
or by an elongated body with 
$n/\Omega=2/4$, $\phi_{2/4}= (4\lambda - 2L' - 2\varpi)/2$.
Right~- 
Phase portrait close to the fourth-order resonance caused by for instance an elongated body with 
$n/\Omega=2/6$, $\phi_{2/6}= (6\lambda - 2L' - 4\varpi)/4$.
}%
\label{chap_rings:fig_phase_portraits}}
\end{figure}
Phase portraits are displayed in \weblink{Fig.}{chap_rings:fig_phase_portraits}, 
using the formalism of \cite{mur99}.
Each panel shows the level curves of the resonant hamiltonian, for a given Jacobi constant. 
Note that while the left panel can only be reproduced in the case of a mass anomaly (1/2 resonance),
the central panel can been either obtained with the 1/3 resonance with a mass anomaly,
or with the 2/4 resonance with an elongated body.
Similarly, the rightmost portrait can be obtained either with a 2/6 resonance with
an elongated body, or with a 1/5 resonance with a mass anomaly.

The topology of each portrait provides interesting hints about the ring response.
Collisions actually tend to damp the forced eccentricities, 
i.e. to attract the eccentricity vector $(X,Y)$ in \weblink{Fig.}{chap_rings:fig_phase_portraits}
towards the origin $X=Y=0$.
Conversely, the resonance tends to force that vector to move around 
a stable fixed point of the phase portrait.

Those antagonist effects may result in an equilibrium.
In the case of a first-order resonance (\weblink{Fig.}{chap_rings:fig_phase_portraits}, left panel),
$(X,Y)$ stabilizes somewhere between the origin and the fixed point, but not along the $OX$-axis.
This creates the torques that are discussed in \weblink{Section}{chap_rings:sec_torques}.
The case of a 2nd-order resonance (middle panel) is different 
because the origin is an unstable hyperbolic point, at least near the resonance. 
As particles are driven towards the origin, the resonance forces it to move
around one of the two stable points, i.e. to acquire large orbital eccentricities.
It is again expected that $(X,Y)$ stabilizes somewhere between the origin and
one of the stable points, but not along the $OY$-axis.
To our knowledge, no work has been done yet to describe the torque that appears in this case.
In the case of a 4th-order resonance (right panel), it can be shown that the origin is always a
stable equilibrium point.
Thus, it is anticipated that the excited streamlines near such resonances damp down to circular orbits,
so that these resonances have eventually little effects on a collisional disk. 
However, this still has to be tested, e.g. using collisional codes, 
and remembering that more subtle effects that are not considered here may show up, 
like viscous effects between neighbor streamlines.

\section{Rings and satellite formation}
\label{chap_rings:sec_sat}

Rings around small bodies are an interesting route towards satellite formation.
The resonant mechanisms described here show that non-axisymmetric
bodies tend to push the initial disk material surrounding the body to more external regions. 
In fact, binarity or multiplicity is common among TNOs \cite{nol08,fra17}.
Although not so common \cite{ric06}, binary asteroids tend to be better known due to their proximity.

\begin{figure}[!t]
\FIG{
\includegraphics[width=0.7\textwidth]{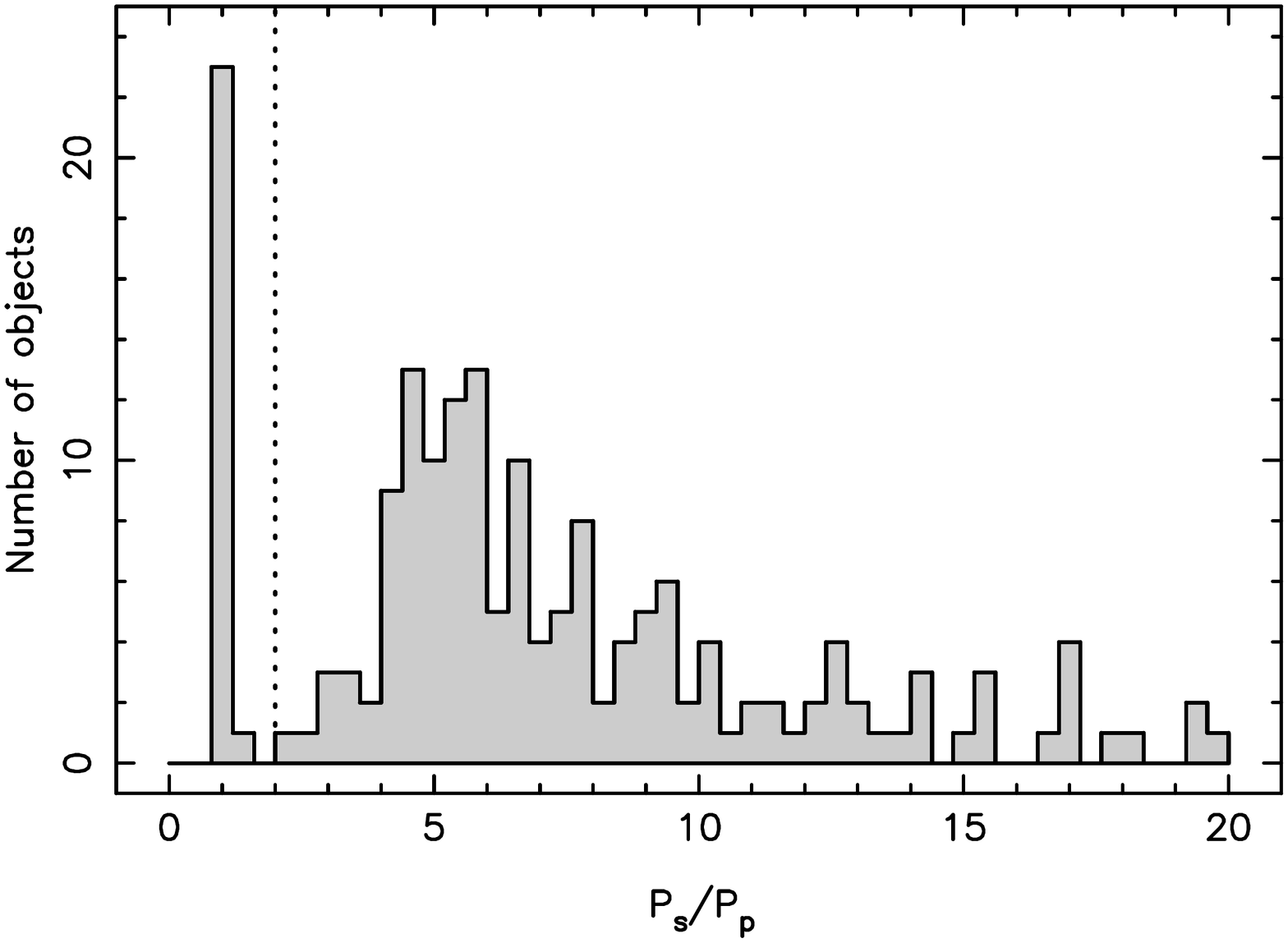}
}
{\caption{%
\title{Orbital periods of satellites compared to spin of primaries.}
The orbital periods $P_{\rm s}$ of 
179 satellites known around multiple asteroids and TNOs
have been normalized to the rotation period $P_{\rm p}$ of their respective primaries
The resulting histogram shows a peak near unity, 
corresponding to tidally evolved synchronous systems.
The 1/2 outer resonance is shown as a dotted line.
The increase of satellite number beyond that resonance
may be a result of the repulsive torques at outer LRs discussed in this chapter.
Taken from \cite{sic18b}.
\label{chap_rings:fig_histo_period_ratio}}
}
\end{figure}

For TNOs, dynamical models are usually invoked to explain 
the formation of binaries with similar masses and wide separation. 
Gravitational interactions between two TNOs can lead to binary formation 
provided that energy is dissipated to bind the pair \cite{nes10}. 
This dissipation could involve dynamical friction with small bodies \cite{gol02}, 
or the intervention of a third body  \cite{sch08}.
For small asteroids (less than 10~km in size) close to the Sun, YORP fission may be a route to satellite formation.
For other asteroids, collisional origin may be a good explanation  \cite{wal15}.

\weblink{Figure}{chap_rings:fig_histo_period_ratio} shows the distribution of 
satellite orbital periods normalized to the primary rotation 
period\footnote{http://www.johnstonsarchive.net/astro/astmoontable.html as of April 2018.}, 
$P_{\rm s}/P_{\rm p}$ (i.e. the inverse of the ratio $n/\Omega$ considered earlier).
Apart from the peak at $P_{\rm s}/P_{\rm p}=1$, due to tidal locking into synchronous orbit,
an interesting feature of the histogram is the absence of satellites with $P_{\rm s}/P_{\rm p}$ between 1 and 2.
This might be the footprint of an ancien collisional disks (whatever their origin)
that suffered a strong coupling with the central body though sectoral resonances,
and thus moved outside the outermost 1/2 resonance to finally accrete into a satellite.
Note finally that in light of our results, 
the formation of satellites outside the 1/2 resonance is favored for slow rotators ($P_{\rm p} > 7$~h),
for which the Roche limit will be inside the aforementioned resonance.

\section{Conclusions}
\label{chap_rings:conclu}

Studies of rings around small bodies are just at their beginning.
Future stellar occultations 
can pin down further the physical parameters of the already known systems
(width, optical depth, orbital parameters) 
and possibly reveal new rings around other Centaurs and TNOs.

However, direct imaging remain difficult, 
owing to the very small angular spans of those rings,  
less than 100~mas for the material detected so far around Chariklo, Haumea and Chiron.
Meanwhile, spectroscopy is challenging too, due to the proximity of the central body
that prevents a clear separation of the ring contribution from the total flux,
although some long-term observations may disentangle to two \cite{duf14}.

Consequently, basic questions such as 
the existence of small moons (shepherding or not the rings),
prograde or retrograde nature of the ring particle motion, or
extended dust sheets around the main rings
remain unanswered.
Future large telescopes, such as the European Extremely Large Telescope (E-ELT)
or the Thirty Meter Telescope (TMT) may have sufficient angular resolution to resolve
these features.

At the other end of methodologies, theoretical approaches will help understand better
the rich dynamical environments of non-axisymmetric bodies. 
In contrast to what happens around giant planets, 
the strong coupling between the spin of the body and the particle mean motions 
is expected to create previously unknown processes in the rings. 
In that context, collisional codes should be most useful to test some of the 
mechanisms that are expected to shape a collisional disk around a small body.
\begin{appendix}


\def\thesection{A.\arabic{section}}		
\def\theequation{A.\arabic{equation}}		
\setcounter{equation}{0}

\section{Appendix}

\subsection{Potential of a homogeneous ellipsoid}
\label{chap_rings:appendix1}

We consider the expansion of the potential created by a homogeneous ellipsoid
of semi-axes $A>B>C$,
see details in \cite{sic18b,bal94,boy97}.
The coefficients $U_{2p}(r)$ in \eqweblink{Eq.}{chap_rings:eq_pot_ell_generic}
can be expanded in power of $R/r$, 
\begin{align}
\label{chap_rings:eq_pot_ell_appen}
\displaystyle
U_{2p} (r) = 
-\frac{GM}{r} 
\sum_{l=|p|}^{+\infty} \left( \frac{R}{r} \right)^{2l} Q_{2l,2|p|}, 
\end{align}
where the reference radius $R$ is given by
\begin{align}
\label{chap_rings:eq_R_appen}
\displaystyle
\frac{3}{R^2}= \frac{1}{A^2} + \frac{1}{B^2} + \frac{1}{C^2},
\end{align}
and
\begin{align}
\label{chap_rings:eq_Q_2l_2p_appen}
Q_{2l,2|p|}=
\frac{3}{2^{l+2|p|} (2l+3)}
\frac{(2l+2|p|)! (2l-2|p|)! l!}{(l+|p|)!(l-|p|)!(2l+1)!} 
\times
\sum_{i=0}^{{\rm int}\left ( \frac{l-|p|}{2} \right )}
\frac{1}{16^i}
\frac{\epsilon^{|p|+2i}}{\left( |p|+i \right)! i!}
\frac{f^{l-|p|-2i}}{\left(l-|p|-2i\right)!}.
\end{align}
The dimensionless parameters $\epsilon$ and $f$
measure the elongation and oblateness of the body, respectively: 
\begin{align}
\label{chap_rings:eq_epsilon_f_appen}
\epsilon = \frac{A^2-B^2}{2R^2} {\rm~~and~~} 
f = \frac{A^2+B^2-2C^2}{4R^2}.
\end{align}
For bodies close to spherical, $(A-B)/R \ll 1$, we retrieve the classical definition of
oblateness, $f \sim (A-C)/A$ and elongation, $\epsilon \sim (A-B)/A$. 

The coefficient $Q_{2l,2|p|}$ is of order $(\epsilon  f)^l$.
For order of magnitude considerations, 
it is enough to consider only the term of lowest order in \eqweblink{Equation}{chap_rings:eq_pot_ell_appen}, i.e. $l=|p|$.
Defining the sequence $S_{|p|} = Q_{2|p|,2|p|}/\epsilon^{|p|}$ and
turning back to $m=2p$, we obtain the expressions in 
\eqweblink{Eqs.}{chap_rings:eq_Sp} and \eqweblink{}{chap_rings:eq_pot_ell_approx}.

\subsection{Mean motion and epicyclic frequency}
\label{chap_rings:appendix2}

The mean motion and epicyclic frequency of a ring particle can be obtained using the classical
formulae
\begin{align}
\label{chap_rings:eq_n_kappa}
n^2(r)= \frac{1}{r} \frac{dU_0(r)}{dr} 
{\rm ~and~} 
\kappa^2(r)= \frac{1}{r^3} \frac{d(r^4 n^2)}{dr},
\end{align}
where $U_0(r)$ is the azimuthally averaged potential.
Using the results above for a homogeneous ellipsoid, and 
keeping the lowest order term in $R/r$, we have
$U_0 (r) \sim -(GM/r)\left[1 + (f/5) (R/r)^2 \right]$, so that
\begin{align}
\label{chap_rings:eq_n_kappa_approx}
n^2(r) \sim \frac{GM}{r^3} \left[1 + \frac{3f}{5} \left(\frac{R}{r}\right)^2\right] 
{\rm ~and~} 
\kappa^2(r) \sim \frac{GM}{r^3} \left[1 - \frac{3f}{5} \left(\frac{R}{r}\right)^2\right].
\end{align}

\subsection{Lindblad Resonance strengths}
\label{chap_rings:appendix3}

The coefficient ${\cal A}_m$ defining the strength of the $m/(m-1)$ LRs are given in \cite{sic18b}.
In the case of the mass anomaly of relative mass $\mu$ sitting at the equator of a sphere of radius $R$, 
it reads (where $a$ is the semi-major axis at the resonance)
\begin{align}
\label{chap_rings:eq_Am_ma}
{\cal A}_m= 
\left\{\left[m+ \frac{a}{2} \frac{d}{da} \right] b^{(m)}_{1/2} \left(\frac{a}{R}\right)+ q \left(\frac{a}{2R}\right) \delta_{(m,-1)} \right\} \mu.
\end{align}
In the case of an elongated body, it is 
\begin{align}
\label{chap_rings:eq_Am_ell}
{\cal A}_m= 
\left[2m-(|m|+1)\right] S_{|m/2|} \left(\frac{R}{a}\right)^{|m|+1} \epsilon^{|m/2|},
\end{align}
where the reference radius is given by \eqweblink{Eq.}{chap_rings:eq_R_appen}.

\end{appendix}

\begin{backmatter}

\begin{ack}[ACKNOWLEDGMENTS]

The work leading to this results has received funding from the 
European Research Council under the European Community's H2020
2014-2020 ERC Grant Agreement n$^\circ$ 669416 ``Lucky Star".
P.S.-S. acknowledges financial support by the European Union's Horizon 2020 Research
and Innovation Programme, under Grant Agreement No. 687378 (SBNAF).

\source{}
\end{ack}


\bibliographystyle{Vancouver-Numbered-Style}
\bibliography{reference}

\end{backmatter}

\end{document}